\documentclass[prd,superscriptaddress,nofootinbib,a4paper,preprintnumbers]{revtex4} 

\usepackage{graphicx}% Include figure files
\usepackage{dcolumn}% Align table columns on decimal point
\usepackage{bm}% bold math
\usepackage{latexsym}
\usepackage{amsfonts}
\usepackage{amssymb}
\usepackage{amsmath}
\usepackage{bbold}
\usepackage[utf8]{inputenc} % set input encoding (not needed with XeLaTeX)
\usepackage{xspace}
\newcommand{\Mpl}{{M_{\rm pl}\xspace}}
\begin{document}

\preprint{Imperial/TP/2017/AEG/1, YITP-17-66, IPMU17-0096}
\title{Stable cosmology in ghost-free quasidilaton theory}
\author{A. Emir G\"umr\"uk\c{c}\"uo\u{g}lu}
\affiliation{Theoretical Physics Group, Blackett Laboratory,
Imperial College London, South Kensington Campus, London, SW7 2AZ,
UK}

\author{Kazuya Koyama}
\affiliation{Institute of Cosmology and Gravitation, University of Portsmouth, Portsmouth PO1 3FX, UK} 

\author{Shinji Mukohyama}
\affiliation{Center for Gravitational Physics, Yukawa Institute for Theoretical Physics, Kyoto University, 606-8502, Kyoto, Japan}
\affiliation{Kavli Institute for the Physics and Mathematics of the Universe (WPI), The University of Tokyo Institutes for Advanced Study, The University of Tokyo, Kashiwa, Chiba 277-8583, Japan}
\begin{abstract}
We present a novel cosmological solution in the framework of extended quasidilaton theory which underwent scrutiny recently. We only consider terms that do not generate the Boulware-Deser degree of freedom, hence the {\it ghost-free} quasidilaton theory, and show three new branches of cosmological evolution therein. One of the solutions passes the perturbative stability tests. This new solution exhibits a late time self-acceleration and all graviton polarizations acquire masses that converge to a constant in the asymptotic future. Moreover, all modes propagate at the speed of light.
We propose that this solution can be used as a benchmark model for future phenomenological studies.
\end{abstract}

\maketitle

\section{Introduction}
The question of a Lorentz invariant finite range gravitational theory has been a long lasting problem of field theory. The first linear proposal of Pauli and Fierz \cite{bib:Pauli-Fierz} has no continuous limit to General Relativity (GR) \cite{bib:ivdvz} unless self-interactions of the graviton are included \cite{Vainshtein:1972sx}. The nonlinear completion of Pauli-Fierz theory was only recently introduced by de Rham, Gabadadze and Tolley (dRGT) \cite{deRham:2010kj} such that the dangerous sixth graviton degree of freedom, dubbed the Boulware-Deser (BD) ghost \cite{Boulware:1973my}, is removed at all orders \cite{Hassan:2011hr}.

The dRGT model forms the basis of contemporary study of Lorentz invariant massive gravity. In addition to the physical metric $g_{\mu\nu}$ which couples directly to matter fields, the construction also needs a nondynamical fiducial metric
\begin{equation}
f_{\mu\nu} = \eta_{ab}\partial_\mu\phi^a\partial_\nu\phi^b\,,
\end{equation}
which is invariant under the Poincar\'e transformations in the internal space of the four St\"uckelberg fields $\phi^a$. The graviton mass is then introduced by tracing various powers of the building block tensor $(\sqrt{g^{-1}f})^\mu_{\;\;\nu}$, consisting of four independent combinations obtained by requiring that the BD mode becomes an auxiliary field at all orders. 

In addition to being a consistent and continuous massive extension of General Relativity, an attractive feature of the theory is the prospect of a self accelerating late time cosmology that obviates the need for a cosmological constant. However, despite the extensive activity in the field, sensible cosmological models within theoretically consistent massive gravity theories pose various challenges for phenomenological studies, due to their complexity. 
Although the original dRGT theory admits open Friedmann-Lema\^itre-Robertson-Walker (FLRW) cosmology \cite{Gumrukcuoglu:2011ew}, it has infinite strongly coupling at linear order \cite{Gumrukcuoglu:2011zh} and exhibits a ghost instability nonlinearly \cite{DeFelice:2012mx}. 
The cosmological backgrounds that have been shown to be perturbatively stable in dRGT theory and its extensions can be classified as follows: i) either homogeneity \cite{D'Amico:2011jj} or isotropy \cite{DeFelice:2013awa} in the fiducial and/or physical metrics is broken; ii) severe fine tunings are necessary \cite{DeFelice:2014nja}; iii) an evolution practically indistinguishable from standard cosmology \cite{Akrami:2015qga}~\footnote{Yet another possibility is to break Lorentz invariance at cosmological scales to realize a nonlinear completion of self-accelerating dRGT cosmology \cite{DeFelice:2015hla,DeFelice:2017wel}. However, we shall not consider this possibility in the present paper.}

The quasidilaton theory was introduced as a simple extension which exploits the stability of Minkowski space-time in dRGT theory by coupling an additional scalar $\sigma$ to the fiducial metric through a conformal transformation, thus sourcing a cosmological expansion \cite{DAmico:2012hia}. The new scalar field, the {\it quasidilaton}, has a restricted action that is invariant under the global transformations
\begin{equation}
\sigma \to \sigma+\sigma_0\,,\qquad
\phi^a\to {\rm e}^{-\sigma_0/\Mpl}\phi^a\,,
\label{eq:globaltransformation}
\end{equation}
that keep the combination ${\rm e}^{2\,\sigma/\Mpl} f_{\mu\nu}$ invariant. The graviton mass term has the same structure as in dRGT with a modified building block ${\rm e}^{\sigma/\Mpl}(\sqrt{g^{-1}f})^\mu_{\;\;\nu}$, while the quasidilaton field acquires its dynamics from the finite number of terms allowed by the symmetry \eqref{eq:globaltransformation}. Although the cosmological solution in this theory suffers from a ghost instability \cite{Gumrukcuoglu:2013nza,DAmico:2013saf}, the coupling between the St\"uckelberg fields and the quasidilaton can extend beyond a conformal factor \cite{DeFelice:2013tsa}, via an extended fiducial metric 
\begin{equation}
\tilde{f}_{\mu\nu} \equiv \eta_{ab}\partial_\mu\phi^a\partial_\nu\phi^b-\frac{\alpha_\sigma}{m^2}\partial_\mu\left({\rm e}^{-\sigma/\Mpl}\right)\partial_\nu\left({\rm e}^{-\sigma/\Mpl}\right)\,,
\label{eq:tildefdefined}
\end{equation}
which under \eqref{eq:globaltransformation}, transforms as
\begin{equation}
\tilde{f}_{\mu\nu} \to {\rm e}^{-2\,\sigma_0/\Mpl} \tilde{f}_{\mu\nu}\,.
\end{equation}
The graviton mass term in the {\it extended quasidilaton theory} can then be constructed using the new building block tensor ${\rm e}^{\sigma/\Mpl}
(\sqrt{\smash[b]{g^{-1}\tilde{f}}})^\mu_{\;\;\nu}$, with additional interactions that are allowed by \eqref{eq:globaltransformation} and that lead to second order equations of motion \cite{DeFelice:2013dua}.

However, recent work demonstrated that the absence of BD ghost is not guaranteed away from the self-accelerating attractor \cite{Anselmi:2017hwr,Golovnev:2017zxk}. This result is compatible with the findings of Ref.~\cite{Kluson:2013jea} and was recently confirmed as a no-go result \cite{Mukohyama:2013raa}. Both analyses indicate that the primary constraint that removes the BD ghost at all orders can be obtained either if $\alpha_\sigma=0$ or if the canonical kinetic term for the quasidilaton field vanishes. Since the former case is simply the original quasidilaton proposal that does not allow for a stable self-accelerating FLRW solution, we will focus on the unexplored latter option. In fact, the absence of explicit kinetic terms for the quasidilaton field does not change its dynamical nature. Instead, the extended fiducial metric readily introduces the dynamical terms for the scalar field. 
The goal of the present study is to focus only on the healthy subset of extended quasidilaton theory in a vacuum configuration. We shall present a new branch of perturbatively stable cosmological evolution, which exhibits a late time self-acceleration with perturbations propagating at the speed of light.

The paper is organized as follows. In Sec.~\ref{sec:theory}, we review the healthy subset of the theory and describe different branches of background evolution. We then follow with Sec.~\ref{sec:perturbations} where we perform a detailed analysis of cosmological perturbations around each of these background solutions, and present stability conditions imposed on the parameters. We conclude with Sec.~\ref{sec:discussion} where we discuss our results.

\section{Ghost-free quasidilaton theory and background dynamics}
\label{sec:theory}

We now consider the action without the canonical kinetic term of quasidilaton, which was shown to be ghost free in Refs.~\cite{Kluson:2013jea,Mukohyama:2013raa}. 
The ghost-free quasidilaton action is thus
\begin{equation}
S=\frac{\Mpl^2}{2}\int d^4 x\sqrt{-g}\left[R+2\,m^2(\alpha_2{\cal L}_2+\alpha_3{\cal L}_3+\alpha_4{\cal L}_4)\right]
\,.
\label{eq:action}
\end{equation}
The mass terms are given by
\footnote{One can also add a {\it cosmological constant} term for the metric ${\rm e}^{2\sigma/\Mpl} \tilde{f}_{\mu\nu}$ which preserves the primary constraint that removes the BD ghost \cite{Mukohyama:2013raa}. As we show in Appendix.\ref{app:xiterm}, this corresponds to simply shifting the parameters of the theory and does not change our results qualitatively.}
\begin{align}
{\cal L}_2&=\frac{1}{2!}\,\left([{\cal K}]^2-[{\cal K}^2]\right)\,,\nonumber\\
{\cal L}_3&=\frac{1}{3!}\,\left([{\cal K}]^3-3\,[{\cal K}][{\cal K}^2]+2[{\cal K}^3]\right)\,,\nonumber\\
{\cal L}_4&=\frac{1}{4!}\,\left([{\cal K}]^4-6\,[{\cal K}]^2[{\cal K}^2]+3[{\cal K}^2]^2+8[{\cal K}][{\cal K}^3]-6[{\cal K}^4]\right)\,,
\label{eq:drgtpotentials}
\end{align}
where
\begin{equation}
{\cal K}^{\mu}_{\;\;\nu} \equiv \delta^\mu_\nu - {\rm e}^{\sigma/\Mpl}\,(\sqrt{g^{-1}\tilde{f}})^\mu_{\;\;\nu}\,,
\end{equation}
and $\tilde{f}_{\mu\nu}$ is defined in Eq.~\eqref{eq:tildefdefined}.
In the present work, our goal is to show the vacuum stability of cosmological solutions with self-acceleration, thus we leave the effect of matter fields for a future work.

We now study the cosmological background in this theory. The St\"uckelberg field background is chosen in order to keep time reparametrisation freedom as
\begin{equation}
\phi^a = a_0 \delta^a_i\,x^i+\delta^a_0\,f(t)
\,,
\end{equation}
and we also have
\begin{equation}
\sigma = \sigma(t)\,.
\end{equation}
These condensates lead to the extended fiducial metric
\begin{equation}
ds_{\tilde{f}}^2 = -n(t)^2\,dt^2+a_0^2\delta_{ij}dx^idx^j\,,
\end{equation}
with an effective lapse function given by
\begin{equation}
n(t)^2 \equiv \dot{f}^2+\frac{\alpha_\sigma}{\Mpl^2m^2}\,{\rm e}^{-2\sigma/\Mpl}\,\dot{\sigma}^2\,.
\label{eq:nlapsedefined}
\end{equation}
Here, an overdot represents derivative w.r.t. the proper time $t$. Finally, we consider a physical metric in flat FLRW form
\begin{equation}
ds^2= -N(t)^2dt^2 +a(t)^2\,\delta_{ij}dx^idx^j\,.
\end{equation}
The resulting minisuperspace action is thus 
\begin{equation}
\frac{S}{V}= \Mpl^2\int dt \,a^3 N \left[\left(-3\,H^2\right)+m^2\Mpl^2(r\,X\,Q-\rho_m)\right]\,,
\label{eq:minisuperspace}
\end{equation}
where we defined
\begin{equation}
H\equiv \frac{\dot{a}}{a\,N}\,,\qquad
X \equiv \frac{a_0{\rm e}^{\sigma/\Mpl}}{a}\,,\qquad
r \equiv \frac{n/a_0}{N/a}\,.
\label{eq:XHrdefined}
\end{equation}
In the above, $H$ is the Hubble rate for the physical metric, $X$ is the ratio of scale factors of the metrics ${\rm e}^{2\,\sigma/\Mpl} \tilde{f}_{\mu\nu}$ and $g_{\mu\nu}$, while $r$ corresponds to the ratio of the speeds of light on these two metrics. For later convenience, we also defined the following useful combinations to replace $\alpha_2$, $\alpha_3$ and $\alpha_4$
\begin{align}
\rho_m &\equiv (X-1)\left[3(2-X)\alpha_2+(X-1)(X-4)\alpha_3+(X-1)^2\alpha_4\right]\,,\nonumber\\
J &\equiv (3-2\,X)\alpha_2+(X-3)(X-1)\alpha_3+(X-1)^2\alpha_4\,,\nonumber\\
Q &\equiv (X-1)\left[3\,\alpha_2-3\,(X-1)\alpha_3+(X-1)^2\alpha_4\right]\,.
\label{eq:RJQdefined}
\end{align}

By varying the minisuperspace action \eqref{eq:minisuperspace} with respect to $N$, $a$, $f$ and $\sigma$, we obtain the following set of background equations of motion
\footnote{We remark that Eqs.\eqref{eq:EQN}--\eqref{eq:EQS} coincide with the ones in \cite{Gumrukcuoglu:2016hic} after the replacement $U'(X)\to -4\,Q$ and $\Omega\to0$.}
\begin{gather}
  \label{eq:EQN}
  3 H^2 = m^2 \rho_m\,,\\
  \label{eq:EQA}
  \frac{2\,\dot{H}}{N}=m^2(r-1)J\,X\,,\\
  \label{eq:EQF}
  \frac{d}{dt}\left[\frac{m^2\Mpl^2a^4Q\,X\,\dot{f}}{n}\right]=0\,,\\
  \label{eq:EQS}
  \frac{\alpha_\sigma}{N\,X\,a^4}\,\frac{d}{dt}\,\left(\frac{a^4\,Q\,\dot{\sigma}}{r\,N\,\Mpl}\right) = m^2 X\left[3\,J(r-1)+4\,Q\,r\right]\,,
\end{gather}
where one of the last three equations is redundant thanks to the contracted Bianchi identities.
Although the form of the St\"uckelberg equation of motion \eqref{eq:EQF} is similar to the well studied ones in e.g. Ref.\cite{DeFelice:2013dua,Gumrukcuoglu:2016hic,Anselmi:2017hwr,Golovnev:2017zxk}, the above system of equations reveals a very useful constraint arising from the absence of a canonical kinetic term for $\sigma$. Combining the time derivative of \eqref{eq:EQN} with \eqref{eq:EQA}, we obtain the simple relation
\begin{equation}
m^2J\,X\,\left(H\,r-\frac{\dot{\sigma}}{\Mpl\,N}\right)=0\,,
\label{eq:constraint}
\end{equation}
where we used that $\dot{\rho}_m = 3\,J\,\dot{X}$ and $ \dot{X}/(N\,X) = \dot{\sigma}/(N\,\Mpl)- H$.

Eq.~\eqref{eq:constraint} is the key difference with respect to the evolution studied in Ref.~\cite{DeFelice:2013dua}, and allows us to determine two branches of evolution without resorting to the integral of Eq.~\eqref{eq:EQF}. These branches are:
\begin{eqnarray}
\frac{\dot{\sigma}}{N\,\Mpl}=& H\,r\,,& \qquad {\rm Branch~I}\,,\nonumber\\
J=&0\,,&\qquad {\rm Branch~II}\,.
\end{eqnarray}
We now consider each branch separately.

\subsection{Branch I}
\label{sec:Branch1}
In this branch, the derivative of the quasidilaton background is solved by 
\begin{equation}
\frac{\dot{\sigma}}{N\,\Mpl}= H\,r\,.
\end{equation}
Using this equation to replace $\dot{\sigma}$ and $\ddot{\sigma}$ in Eq.~\eqref{eq:EQS}, then eliminating $\dot{H}$ using Eq.~\eqref{eq:EQA}, we find
\begin{align}
2\,\left(\alpha_\sigma H^2-m^2X^2\right)\left[4\,Q+3(r-1)(J+Q)\right]
+m^2Q\,X(r-1)(\alpha_\sigma J-2\,X)=0\,,
\label{eq:eqr}
\end{align}
which can be used to express $r$ in terms of $X$.~\footnote{More precisely, we assume here that the coefficient of $r$ is nonzero. However, the special cases where $r$ cannot be determined from Eq.\eqref{eq:eqr} all lead to a constant $X$, therefore to $r=1$. As a result, all of these cases belong to one of the late time evolutions given in Eq.\eqref{eq:BranchIlate}.}
Thus with the help of Eq.\eqref{eq:EQN}, all cosmological quantities can now be written in terms of $X$. By further manipulating the equations, one can also determine the evolution equation for $X$,
\begin{equation}
\frac{\dot{a}}{a} = -\frac{3}{8}\left[
\frac{2\,(J+Q)}{Q\,X}+\frac{\alpha_\sigma J-2\,X}{\alpha_\sigma \rho_m-3\, X^2 }
\right]\dot{X}\,.
\end{equation}
This equation can be integrated to give the following relation:
\begin{equation}
Q^2\left(\alpha_\sigma \rho_m-3\,X^2 \right) = \frac{\lambda}{a^8}\,,
\label{eq:Asolution}
\end{equation}
where $\lambda$ is an integration constant. 

We now discuss the late time asymptotics $a\to\infty$. In this regime, we expect the left hand side of Eq.~\eqref{eq:Asolution} to approach zero. The two possibilities are 
\begin{eqnarray}
H\to&\frac{m\,X}{\sqrt{\alpha_\sigma}} \,,& \qquad {\rm Branch~IA}\,,\nonumber\\
Q\to&0\,,&\qquad {\rm Branch~IB}\,.
\label{eq:BranchIlate}
\end{eqnarray}
By comparing with \eqref{eq:EQS} and using the definitions in \eqref{eq:nlapsedefined} and \eqref{eq:XHrdefined}, we see that for Branch--IA, one has $\dot{f}\to0$, while Branch-IB is the $Q\to0$ branch discussed in \cite{DeFelice:2013dua}.\footnote{
It should be noted that the general study in Ref.~\cite{DeFelice:2013dua} cannot be directly used to obtain the results of Branch--IB of the present paper by simply sending the coefficient of the quasidilaton kinetic term to zero. Firstly, these studies use the background equations by assuming that the kinetic term is nonzero. Also due to our late time behaviour of $r\to 1$, the perturbative stability conditions occasionally depend on factors of $0/0$, thus we repeat the analysis for Branch--IB without referring to previous work in more general backgrounds.}
 In Branch-IB, we also find an important condition on background variables in the late asymptotics, namely,
\begin{equation}
 \frac{\dot{f}^2}{N^2} \to {\rm e}^{-2\,\sigma/\Mpl}\,\left(X^2-\frac{\alpha_\sigma H^2}{m^2}\right) > 0 \,,\qquad {\rm Branch~IB}\,,
 \label{eq:IBcondition-late}
\end{equation}
since the background values of the St\"uckelberg fields are real by definition. We will see that this condition plays a very crucial role in the stability of scalar perturbations in this branch.

To collect, the various limits in the two sub-branches are
\begin{eqnarray}
X \to \frac{\sqrt{\alpha_\sigma} H}{m}\,,\qquad
\dot{H}\to 0\,,\qquad
\dot{X}\to 0\,,\qquad
r\to 1\,,\qquad
\dot{\sigma}\to \Mpl\,N\,H\,,& \qquad {\rm Branch~IA}\nonumber\\
Q\to 0\,,\qquad
\dot{H}\to 0 \,,\qquad
\dot{X}\to 0\,,\qquad
r\to 1\,,\qquad
\dot{\sigma}\to \Mpl\,N\,H\,,& \qquad {\rm Branch~IB}.
\label{eq:branchI-asymptotics}
\end{eqnarray}

The main difference between the two sub-branches is that the fixed point value of $X$ is determined by the equation $(\alpha_\sigma\rho_m-3\,X^2)=0$ for IA, whereas for IB it is simply $Q=0$ (3 solutions in both cases).

The Branch--IA corresponds to the $\dot{f}\to0$ attractor first introduced in Ref.~\cite{Gumrukcuoglu:2016hic}, and later studied in detail in Ref.~\cite{Anselmi:2017hwr} in the context of extended quasidilaton. Due to the absence of a canonical kinetic term for the quasidilaton, the $\dot{f}\to 0$ case actually corresponds to two distinct branches of the evolution in the present study, the other being the Branch--II which we consider next.

\subsection{Branch II}
\label{sec:Branch2}
Since the quasidilaton kinetic term does not contribute to the metric equations, using $J=0$ in Eq.~\eqref{eq:EQA} reveals that this branch of evolution is purely de Sitter. The constancy of $X$ results in $\dot{\sigma} = \Mpl H\,N$, while the quasidilaton equation of motion yields a first order equation for $r$, which for $Q\neq0$, reads
\begin{equation}
\frac{\dot{r}}{N}-4\,H\,r+\frac{4\,m^2\,r^3\,X^2}{\alpha_\sigma H}=0\,.
\end{equation}
The solution to this equation yields the evolution of $r$
\begin{equation}
r = \frac{a^4\,H\,\sqrt{\alpha_\sigma}}{\sqrt{C_1+a^8 m^2 X^2}}\,.
\end{equation}
We thus see that in the late time asymptotics, $r\to H\sqrt{\alpha_\sigma}/(m\,X)$. To recap, the late time behaviour for this branch is given by
\begin{equation}
J\to 0\,,\qquad
\dot{H}\to 0 \,,\qquad
\dot{X}\to 0\,,\qquad
r\to \frac{\sqrt{\alpha_\sigma}H}{m\,X}\,,\qquad
\dot{\sigma}\to \Mpl\,N\,H\,, \qquad {\rm Branch~II}.
\end{equation}
This branch, along with Branch--IA, corresponds to the $\dot{f}\to0$ branch introduced in Ref.~\cite{Gumrukcuoglu:2016hic}. However, as we will show when we discuss cosmological perturbations, the kinetic terms of three graviton polarizations are proportional to $J$ and thus the absence of the quasidilaton kinetic term has a dramatic impact on the validity of the perturbative expansion. Notice that in the vacuum case, $J=0$ is not an asymptotic behaviour, but enforced by the equations of motion at all times.  

\section{Cosmological Perturbations}
\label{sec:perturbations}
We now introduce perturbations to the metric and the five scalar fields. We decompose the metric perturbations as
\begin{align}
 \delta g_{00} &= -2\,N^2\,\Phi\,,\nonumber\\
 \delta g_{0i} &= N\,a\,(\partial_i B + B_i)\,,\nonumber\\
 \delta g_{ij} &= a^2\left[2\,\delta_{ij}\psi + \left(\partial_i\partial_j-\frac{\delta_{ij}}{3}\partial^2\right)E+\partial_{(i}E_{j)}+\gamma_{ij}\right]\,,
 \label{eq:decomposition}
\end{align}
while the perturbations to the quasidilaton field is introduced as $\Mpl \delta\sigma$, and the St\"uckelberg field perturbations are
\begin{equation}
\delta\phi^0 = \Pi^0\,,\qquad
\delta\phi^i = \Pi^i+\partial^i \Pi_L\,.
\label{eq:stuckelbergpert}
\end{equation}
In these decompositions, all 3--vectors are divergence-free and the 3--tensor $\gamma_{ij}$ is both divergence and trace-free with respect to $\delta^{ij}$. We choose a gauge that is relevant for our background evolution, namely, $E_i = E = \delta\sigma=0$, which fixes the freedom completely even in the case $\dot{f}\to0$ as shown in Appendix \ref{sec:gaugechoice}.

In the following, we consider each sector independently. 

\subsection{Tensor Sector}
After expanding the fields in terms of plane waves, adding appropriate boundary terms and using background equations of motion, we obtain the following action for tensor perturbations
\begin{equation}
S_T^{(2)} = \frac{\Mpl^2}{8}\int d^3k\,dt\, a^3 N\,\left(
\frac{\dot{\gamma}_{ij}^\star}{N}\frac{\dot{\gamma}^{ij}}{N}-\omega_{\rm T}^2 \gamma_{ij}^\star\gamma^{ij}
\right)\,,
\end{equation}
where the kinetic term is manifestly positive and the dispersion relation is given by
\begin{equation}
\omega_{\rm T}^2 = \frac{k^2}{a^2}+M_{T}^2\,,
\end{equation}
with an effective mass 
\begin{equation}
M_T^2 = \frac{m^2\,X}{X-1}
\left[J[r-2+(2\,r-1)X] +\frac{(r-1)[Q-X^2\rho_m]}{(X-1)^2}\right]\,.
\label{eq:masstensor}
\end{equation}

Up to this point, we have not yet specified any branch of evolution. For the three branches discussed in Section \ref{sec:theory}, the tensor mass converges to the following constant values asymptotically
\begin{align}
M_{T,IA}^2 &\to m^2 J\,X\,,\\
M_{T,IB}^2 & \to m^2 J\,X\,,\\
M_{T,II}^2 &\to \frac{m(\sqrt{\alpha_\sigma}H-m\,X)}{(X-1)^3}\left[Q-\rho_m X^2\right]\,.
\label{eq:MTII}
\end{align}
In order to avoid tachyonic instability, we need to have $M_T^2 >0$. This is the condition for the IR stability and thus can in principle be violated without fatal consequences as far as the tensor sector is concerned. However, we shall see below that this condition is required by the UV stability of the vector sector for the Branches IA and IB. We shall also see that the branch II is strongly coupled in the vector and scalar sectors. Therefore, the tensor sector is free from instabilities both in the UV and in the IR whenever the vector and scalar sectors are weakly coupled and stable in the UV. 

\subsection{Vector Sector}
The action quadratic in vector modes does not contain any dynamical term for the $B_i$ mode, whose equation of motion can be solved by
\begin{equation}
B_i = \frac{2\,m^2a^2J\,{\rm e}^{\sigma/\Mpl}}{N[k^2(r+1)+2\,m^2a^2J\,X]}\,\dot{\Pi}_i\,.
\end{equation}
Integrating it out and adding boundary terms, the quadratic action for vector perturbations reduces to
\begin{equation}
S_V^{(2)} = \frac{\Mpl^2}{4}\int d^3k\,dt\, a^3 N\,\frac{k^2}{a_0^2}\left(
{K}_V \,\frac{\dot{\Pi}_i^\star}{N}\frac{\dot{\Pi}^i}{N}-M_T^2\Pi_i^\star \Pi^i
\right)\,,
\end{equation}
where the kinetic term is
\begin{equation}
K_V\equiv \left(\frac{k^2(r+1)}{2\,m^2 a^2 J X}+1\right)^{-1}\,.
\end{equation}
For the three branches of evolution discussed in Sec.\ref{sec:theory}, it reduces to
\begin{align}
K_{V,IA}^2 &= K_{V,IB}^2 =
\left(1+\frac{k^2}{m^2a^2J\,X}\right)^{-1}
\,,\\
K_{V,II}^2 &= 0\,.
\end{align}
We immediately see that in Branch-II, the vacuum evolution strictly requires $J=0$, thus the vector modes become infinitely strongly coupled. 

However, for both the Branch-I evolutions, the UV behaviour of the kinetic term imposes the stability condition
\begin{equation}
J>0\,,\qquad {\rm Branch~I}\,,
\end{equation}
which also coincides with the condition obtained from the IR stability of tensor modes in the previous subsection. By rescaling the field to acquire canonical normalisation, we obtain the following action for Branch--I, in the late time limit
\begin{equation}
S_{V,I}^{(2)} = \frac{\Mpl^2}{2}\int d^3k\,dt\, a^3 N\,\left(\frac{\dot{\Pi}_{c,i}^\star}{N}\frac{\dot{\Pi}^i_c}{N}-\omega^2_{V,I}\Pi_{c,i}^\star \Pi^i_c
\right)\,,
\end{equation}
where $\Pi^i_c$ is the rescaled field and the corresponding dispersion relation is 
\begin{equation}
\omega^2_{V,I} = \frac{k^2}{a^2}+m^2J\,X - \frac{k^2H^2(4\,k^2+m^2a^2J\,X)}{(k^2+m^2a^2J\,X)^2}\,,\qquad{\rm Branches~IA/B,~late~time}\,.
\end{equation}
The first term dominates in the UV, while the second term dominates in the IR. Thus we see that the sound-speed is $1$ and the mass is $m^2J\,X$.

\subsection{Scalar Sector}
Finally, we consider the scalar sector. The two nondynamical fields are solved as

\begin{align}
\Phi &= \frac{1}{N}\partial_t\left(\frac{\psi}{H}\right)+\frac{k^2m^2J}{H\,N\,[2\,k^2(r+1)+3\,m^2a^2J\,X]}\left[
\frac{X(2\,k^2\,\Mpl N\,r^2+3\,m^2a^2J\,X\,\dot\sigma)}{2\,\Mpl\,H^2k^2}\,\psi
+a\,{\rm e}^{\sigma/\Mpl}\dot{\Pi}^L-\frac{{\rm e}^{2\,\sigma/\Mpl}\dot{f}}{X}\,\Pi^0
\right.\nonumber\\
&\left.\qquad\qquad\qquad\qquad\qquad\qquad\qquad\qquad\qquad\qquad\qquad\qquad\qquad\qquad-
\frac{m^2{\rm e}^{\sigma/\Mpl}a\,J\,X\,(\Mpl H\,N(r-1)-\dot\sigma)}{2\,\Mpl H^2}\,\Pi^L
\right]\,,\nonumber\\
B &= -\frac{\psi}{a\,H}+\frac{m^2J}{H\,N\,[2\,k^2(r+1)+3\,m^2a^2J\,X]}\left[
\frac{3\,a\,X[\Mpl H\,N\,r^2-(r+1)\dot\sigma]}{\Mpl\,H}\,\psi
+3\,a^2H\,{\rm e}^{\sigma/\Mpl}\dot{\Pi}^L-\frac{3\,a\,H\,{\rm e}^{2\,\sigma/\Mpl}\dot{f}}{X}\,\Pi^0
\right.\nonumber\\
&\left.\qquad\qquad\qquad\qquad\qquad\qquad\qquad\qquad\qquad\qquad\qquad\qquad\qquad\qquad\qquad\qquad\qquad-
N\,{\rm e}^{\sigma/\Mpl}k^2(r+1)\,\Pi^L
\right]\,,
\end{align}
After using these solutions, the action quadratic in scalar perturbations consists of three modes, $\psi$, $\Pi^0$ and $\Pi^L$. Without any field redefinition, the perturbation $\psi$ has no kinetic term, thus this is  identified as the Boulware-Deser degree. After integrating it out, the two field action can be brought to the form
\begin{equation}
S_S^{(2)}=\frac{\Mpl^2}{2}\int d^3k dt \,a^3 N
\left(\frac{\dot{\Psi}^\dagger}{N}K\frac{\dot{\Psi}}{N}+\frac{\dot{\Psi}^\dagger}{N}M\Psi-\Psi^\dagger M \frac{\dot{\Psi}}{N}-\Psi^\dagger \Omega^2\Psi\right)\,,
\label{eq:scalaractionformal}
\end{equation}
where $\Psi\equiv (\Pi^0,\,\Pi^L)$ and the $2\times2$ matrices obey $K=K^T$, $M=-M^T$, $\Omega^2=(\Omega^2)^T$.

The conditions for having positive kinetic terms for the eigenmodes can be obtained by studying $\kappa_1 \equiv K_{22}$ and $\kappa_2\equiv \det K/K_{22}$ in the UV limit. 

In the following, instead of presenting the rather bulky expressions for the exact matrix components, we will only present them when they are informative. 

\subsubsection{Branch IA}
Let us first consider the Branch--I before taking the late time limit, so that we can use it in Branch--IB. The two kinetic eigenvalues in the subhorizon limit are
\begin{align}
\kappa_{1,I} &= 
\frac{3\,{\rm e}^{2\,\sigma/\Mpl}m^2a^2J(m^2J\,X-2\,H^2)}{2\,X}\left(1+\frac{8\,Q(\alpha_\sigma H^2-m^2X^2)}{2\,\alpha_\sigma H^2(3\,J-Q)+m^2 J\,X(\alpha_\sigma Q-6\,X)}\right)+{\cal O}(k^{-2})
\,,\nonumber\\
\kappa_{2,I} &= \frac{{\rm e}^{2\,\sigma/\Mpl}H^2}{X^3}\left(\alpha_\sigma Q - \frac{2\,(3\,J+4\,Q)(\alpha_\sigma H^2 -m^2X^2)}{(2\,H^2-m^2J\,X)}\right)\,.
\label{eq:branchIkin}
\end{align}

However, once the late time limits \eqref{eq:branchI-asymptotics} for Branch--IA are used, the two scalar modes decouple and the (now diagonal) matrices in the action \eqref{eq:scalaractionformal} reduce to
\begin{eqnarray}
K_{11} &=\frac{a^2m^2Q\,X}{a_0^2}\,,\qquad
K_{22} &= \frac{6\,m^4a^4J\,X^3}{a_0^2\alpha_\sigma {\cal D}}\left(\frac{\alpha_\sigma J}{X}-2\right)\,,\nonumber\\
{\Omega^2}_{11} &= \frac{k^2m^2Q\,X}{a_0^2}\,,\qquad
{\Omega^2}_{22} &= \frac{2\,X\,m^2J\,k^4}{3\,a_0^2}\left\{1-\frac{1}{{\cal D}}\left[4+\frac{3\,a^2H^2}{k^2}\left(\frac{\alpha_\sigma J}{X}-2\right)\left(1-\frac{48\,a^2H^2}{k^2{\cal D}}\right)\right]\right\}\,,
\end{eqnarray}
where
\begin{equation}
{\cal D}\equiv \left[4+\frac{3\,a^4H^4}{k^4}\,\left(\frac{\alpha_\sigma J}{X}-2\right)\left(\frac{4\,k^2}{a^2H^2}+\frac{3\,\alpha_\sigma J}{X}\right)\right]\,.
\end{equation}
The positivity of the kinetic terms imposes the following conditions:
\begin{equation}
J\left(J-\frac{2\,X}{\alpha_\sigma}\right)>0\,,\qquad
Q>0\,,\qquad {\rm Branch~IA}\,.
\end{equation}
Assuming that these conditions are satisfied, we can rescale the fields and rewrite the action as
\begin{equation}
S_{S,IA}^{(2)}=\frac{\Mpl^2}{2}\int d^3k dt \,a^3 N
\left(\frac{\dot{\Psi}_c^\dagger}{N}\frac{\dot{\Psi}_c}{N}-\Psi_c^\dagger \omega^2_{S}\Psi_c\right)\,,
\label{eq:scalaractionrescaled}
\end{equation}
where $\Psi_c$ are the fields in this normalisation and the dispersion relation $\omega^2_S$ is a diagonal matrix with components
\begin{align}
\omega_{S,1}^2 &= \frac{k^2}{a^2}-4\,H^2\,,\nonumber\\
\omega_{S,2}^2 &= \frac{k^2}{a^2} + m^2J\,X -\frac{12\,H^2}{{\cal D}}\left[2+\frac{a^2H^2}{k^2}\left(\frac{\alpha_\sigma J}{X}-2\right)\left(1-\frac{72\,a^2\,H^2}{k^2{\cal D}^2}\right)\right]\,,\qquad {\rm Branch~IA}.
\end{align}
We see that the first mode propagates at the speed of light and has a negative squared-mass. Although this may indicate some tachyonic instability, contributions which are factors of $H^2$ depends on the normalisation of the field. However, regardless of what the right normalisation is, 
%dictated by the observables, 
the time scale of the corresponding tachyonic instability is of the same order as the age of the universe. Therefore, even if it is present, it is a mild instability. 

For the second mode, the last term is subdominant both on sub- and super-horizon scales, so we see that the mode propagates with the sound speed $c_s^2=1$ and has a mass $m^2J\,X$, just like the vector and tensor perturbations.

\subsubsection{Branch IB}
For this branch, the subhorizon limit expressions for the kinetic terms in \eqref{eq:branchIkin} are still valid and applying the asymptotic $Q\to0$ limit, we find
\begin{equation}
\kappa_{1,IB} \to \frac{3\,{\rm e}^{2\,\sigma/\Mpl}m^2a^2J\,(m^2J\,X-2\,H^2)}{2\,X}\,,\qquad
\kappa_{2,IB} \to -\frac{6\,{\rm e}^{2\,\sigma/\Mpl}m^2H^2J}{X^3(m^2J\,X-2\,H^2)}\,\left(X^2-\frac{H^2\,\alpha_\sigma}{m^2}\right)\,.
\end{equation}
Using \eqref{eq:IBcondition-late}, we see that the product of these two eigenvalues is a negative total square, hence we conclude that this branch always contains a scalar ghost.

\subsubsection{Branch II}
In this branch, $J=0$ is enforced by the (vacuum) equations of motion, and for this value, $\kappa_2$ vanishes. Since the vanishing of $J$ in this branch is not an asymptotic behaviour but rather a dynamical requirement throughout the evolution, the mode is infinitely strongly coupled. Since the vector modes in this branch are also strongly coupled, this is reminiscent of the self-accelerating branch in dRGT massive gravity.

\section{Discussion}
\label{sec:discussion}

We have presented a new cosmological vacuum solution in the context of ghost-free quasidilaton theory. This is a restricted version of the extended quasidilaton theory where all terms that generate a Boulware-Deser ghost are turned off. We have shown that there exists a branch of the cosmological evolution (Branch--IA), which has a self-accelerating solution where all five degrees of freedom have equal masses $\propto m$ and propagate at the speed of light. The additional degree of freedom, identified as the quasidilaton perturbation, has a mass $\propto H$, which may or may not lead to a mild tachyonic instability in the IR, depending on the normalisation imposed by observables.

This solution is, to our knowledge, the simplest model in Lorentz invariant massive gravity that passes the perturbative stability tests while exhibiting self-acceleration. It is thus a promising candidate to be used as a benchmark for phenomenological studies, after its interactions with matter fields is understood in detail. This point will be addressed in a future publication.

Fixing the value of $\alpha_2 m^2$ by using the present day Hubble rate, our model can be reduced to contain three independent parameters $\alpha_3,\alpha_4$ and  $\alpha_\sigma$. The parameter $\alpha_\sigma$ should be strictly positive in order to realise the stable branch of evolution. This is also the parameter that allows the theory to depart from the standard quasidilaton of Ref.\cite{DAmico:2012hia}.

In our cosmological solution, the background value of the temporal St\"uckelberg field flows to a constant at late times. The background evolution does not actually require the time derivative to be strictly zero; instead the VEV of $\dot{\phi}^0$ smoothly approaches to zero as $a^{-5}$ without becoming zero in finite time. In other words, the whole cosmological history in our solution covers only a half of the moduli space. This is similar to the self-accelerating open FLRW solution in dRGT theory, where only a quarter of the moduli space (a Milne wedge of the Minkowski moduli space) is covered by the whole cosmological history~\cite{Gumrukcuoglu:2011ew}. In both cases, the fiducial metric is locally Minkowski and the theory respects the Poincare symmetry in the moduli space. One might still worry that the transformation that takes Minkowski to the fiducial metric in our solution would be singular and one would no longer have a fixed reference metric~\cite{Golovnev:2017zxk}. Actually, we find that this is not a problem. As is clear from the fact that the part (either a half or a quarter) of the moduli space covers the whole cosmological history, one never reaches the boundary at finite time and therefore the transformation to the explicitly Minkowski form of the fiducial metric remains regular. 

As a result of the behaviour $\dot{\phi}^0\to 0$, the unitary gauge $\delta\phi^0=0$ cannot be fixed in the asymptotic future, and we resort to another gauge choice.
Taking the theory in the covariant formulation, $\dot{\phi}^0\to0$ is an allowed background which prefers a certain gauge over the unitary one. 
Despite the smallness of $\dot{\phi^0}$, the perturbative expansion is not affected by potentially vanishing coefficients. In the late time limit of Branch--IA, the tensor $g^{-1}\tilde{f}$ can be expanded as
\begin{equation}
g^{\mu\alpha}\tilde{f}_{\alpha_\nu} = \frac{a_0^2}{a^2}\delta^\mu_\nu +a_0\left(\delta^\mu_j\partial_\nu\delta\phi^j+\delta_\nu^i\eta_{ai}g^{\mu\alpha}\partial_{\alpha}\delta\phi^a\right)
+g^{\mu\alpha}\eta_{ab}\partial_\alpha\delta\phi^a\partial_\nu\delta\phi^b\,,
\end{equation}
where we only kept the St\"uckelberg perturbations $\delta\phi^a$, ($a=0,1,2,3$) for the present discussion. Thus, even for $\dot{\phi}^0=\dot{f}=0$, all the four eigenvalues of $g^{-1}\tilde{f}$ have nonzero VEV's and the time derivatives of the St\"uckelberg perturbations have finite coefficients. This allows the square-root of the tensor to be Taylor expanded up to any order in perturbations without any ambiguity and we do not observe any signs of a strong coupling in this limit. 

In addition to the healthy branch, we found two more late time solutions which do not pass the stability tests: Branch--IB contains a scalar ghost, while in Branch--II two vector and one scalar degree of freedom is infinitely strongly coupled throughout the evolution.

Although the well-being of the theory relies on the absence of canonical kinetic terms for the quasidilaton, there may be other terms that do not excite the Boulware-Deser mode. The setup we considered is reminiscent of DBI scalars coupled to massive gravity \cite{Goon:2011uw,Goon:2011qf,Gabadadze:2012tr}. For the ghost-free quasidilaton, the 3-brane with metric $\tilde{f}_{\mu\nu}$ can be interpreted as probing a bulk with two time directions. This interpretation indicates the existence of further ghost-free terms involving the quasidilaton, corresponding to curvature invariants and Gibbons-Hawking-York boundary terms from 5D Lovelock terms \cite{deRham:2010eu} (see e.g. \cite{Goon:2014ywa} for the cosmology of such a model with a different bulk metric signature). 

\acknowledgments
This work was completed during SM's visit to Imperial College London. He would like to thank Claudia de Rham and Andrew Tolley for their
warm hospitality. AEG acknowledges support by the UK Science and Technologies Facilities Council (STFC) grant ST/L00044X/1. KK is supported by the STFC grant ST/N000668/1. The work of KK has received funding from the European Research Council (ERC) under the European Union's Horizon 2020 research and innovation programme (grant agreement 646702 "CosTesGrav"). 
The work of SM was supported by Japan Society for the Promotion of Science (JSPS) Grants-in-Aid for Scientific Research (KAKENHI) No. 24540256, No. 17H02890, No. 17H06359, No. 17H06357, and by World Premier International Research Center Initiative (WPI), MEXT, Japan.

\appendix
\section{The effect of cosmological constant for the extended fiducial metric}
\label{app:xiterm}
The quasidilaton symmetries \eqref{eq:globaltransformation} allow a cosmological constant term for the extended fiducial metric, in the form
\begin{equation}
\Delta S = \Mpl^2 m^2 \xi\,\int d^4x \,{\rm e}^{4\,\sigma/\Mpl}\sqrt{-\tilde{f}}\,.
\label{eq:deltaS}
\end{equation}
We can express this action purely in terms of dRGT mass terms. We first add a term
\begin{equation}
{\cal L}_1\equiv [{\cal K}]\,,
\end{equation}
to the dRGT mass terms given in \eqref{eq:drgtpotentials}. Including the identity, ${\cal L}_1$, ${\cal L}_2$, ${\cal L}_3$ and ${\cal L}_4$ form the characteristic polynomials for the matrix ${\cal K}$. We can thus perform the expansion
\begin{equation}
\det (I_4-{\cal K}) = 1 -{\cal L}_1 + {\cal L}_2 -{\cal L}_3 +{\cal L}_4\,,
\end{equation}
where $I_4$ is the four dimensional identity matrix. As the left hand side is simply $\det({\rm e}^{\sigma/\Mpl}\sqrt{g^{-1}f})$, we obtain
\begin{equation}
\Delta S = \Mpl^2m^2 \xi\,\int d^4x\,\sqrt{-g}\left[1 -{\cal L}_1 + {\cal L}_2 -{\cal L}_3 +{\cal L}_4\right]\,.
\end{equation}
In other words, the additional term \eqref{eq:deltaS} can be written as a combination of the dRGT mass terms, and corresponds to a shift in the coupling constants.  On the other hand, adding the $\xi$ term would correspond to including the tadpole term ${\cal L}_1$ and a bare cosmological constant, which were not considered in the main text. The effect of $\xi$ can be accommodated by changing the definition of $Q$ in Eq.~\eqref{eq:RJQdefined} to
\begin{equation}
Q \equiv (X-1)\left[3\,\alpha_2-3\,(X-1)\alpha_3+(X-1)^2\alpha_4\right]+\xi\,X^3\,,
\label{eq:Qdefshifted}
\end{equation}
then shifting $Q\to Q -\xi X^3$ in Eqs.~\eqref{eq:masstensor} and \eqref{eq:MTII}. All other occurrences of $Q$ in the text should then be understood to correspond to the definition in Eq.~\eqref{eq:Qdefshifted}.

\section{The gauge choice}
\label{sec:gaugechoice}
As we described in the main text, two out of the three branches of late time evolution require $f'(t)\to0$, namely the background value of the temporal St\"uckelberg field approaches a constant. As a result, its perturbations become effectively gauge invariant, hence the gauge freedom that allows us to set it to zero (i.e. the unitary gauge) is no longer allowed. In order to find a suitable gauge that is valid for all three branches of the evolution considered in the main text, we study the behaviour of the perturbations under linear gauge transformations. 

For a coordinate transformation $x^\mu\to x^\mu+\xi^\mu$, with the decomposition $\xi^\mu = (\xi^0, \xi^i + \partial^i \xi_L)$, the  perturbations \eqref{eq:decomposition}-\eqref{eq:stuckelbergpert} transform via
\begin{align}
\Phi &\to \Phi -\frac{1}{N}\partial_t(N\,\xi^0)\,,\nonumber\\
B &\to B + \frac{N}{a}\,\xi^0 - \frac{a}{N}\,\dot{\xi}_L\,,\nonumber\\
B_i&\to B_i -\frac{a}{N}\dot{\xi}_i\,,\nonumber\\
\psi &\to \psi -\frac{\dot{a}}{a}\,\xi^0-\frac{1}{3}\partial^i\partial_i\xi_L\,,\nonumber\\
E&\to E-2\,\xi_L\,,\nonumber\\
E_i&\to E_i - 2\xi_i\,,\nonumber\\
\delta\sigma &\to \delta\sigma - \frac{\dot\sigma}{\Mpl}\,\xi^0\nonumber\\
\Pi^0 &\to \Pi^0 -\dot{f}\,\xi^0\,,\nonumber\\
\Pi_L &\to \Pi_L -a_0\xi_L\,,\nonumber\\
\Pi^i &\to \Pi^i -a_0\xi^i\,,
\end{align}
and the tensor modes $\gamma_{ij}$ are invariant at linear order. Out of the total of 7 scalar degrees of freedom, one can construct 5 independent gauge-invariant scalar degrees, while the 3 transverse vectors can be reduced to 2 gauge invariant divergence-free vector fields. This choice however is not unique and the specific choice of gauge invariant variables may be equivalent to a a particular gauge choice. For instance, the choice of the gauge invariant variable $\delta\sigma-\dot\sigma\Pi^0/(\dot{f}\Mpl)$ implicitly implies the gauge $\Pi^0=0$. On the other hand, if $\dot{f}=0$ is allowed by the evolution (which is the case for the model studied in the main text), this choice is no longer well defined, nor gauge invariant, and $\Pi^0$ itself becomes gauge invariant on its own. Therefore, we need to construct the set of variables carefully. 

For the transverse vectors, we define the following combinations:
\begin{align}
B^i_{GI} &\equiv B^i - \frac{a}{2N}\,\dot{E}^i\,,\nonumber\\
\Pi^i_{GI} &\equiv \Pi^i - \frac{a_0}{2}\,E^i\,,
\end{align}
and for the scalars, we choose
\begin{align}
\Phi_{GI} &= \Phi-\frac{1}{N}\,\partial_t\left[\frac{1}{H}\left(\psi+\frac{k^2}{6}E\right)\right]\,,\nonumber\\
B_{GI} &= B+\frac{1}{a\,H}\,\left(\psi+\frac{k^2}{6}E\right) - \frac{a}{2\,N}\,\dot{E}\,,\nonumber\\
\psi_{GI} &= \psi +\frac{k^2}{6}\,E-\frac{H\,N\,\Mpl}{\dot{\sigma}}\,\delta\sigma\,,\nonumber\\
\Pi^0_{GI} &= \Pi^0-\frac{\Mpl\,\dot{f}}{\dot\sigma}\,\delta\sigma\,,\nonumber\\
\Pi^L_{GI} &= \Pi^L - \frac{a_0}{2}\,E\,.
\end{align}
We see that these definitions are not affected by the evolution, provided that $\dot\sigma\neq0$. Upon integration of the nondynamical modes $\Phi$ and $B$ (or equivalently, $\Phi_{GI}$ and $B_{GI}$) this choice of variables actually corresponds to the gauge choice $E^i = E=\delta\sigma=0$, which completely fixes the freedom if $\dot\sigma\neq0$.

\end{document}